%Paper: hep-ph/9308287
%From: luke@yukawa.ucsd.edu
%Date: Tue, 17 Aug 93 13:41:31 -0700
%Date (revised): Tue, 17 Aug 93 15:48:44 -0700
%Date (revised): Mon, 1 Nov 1993 17:26:26 -0800

\input epsf

\ifx\epsffile\undefined\message{(FIGURES WILL BE IGNORED)}
\def\insertfig#1#2{}% null macro
\else\message{(FIGURES WILL BE INCLUDED)}
\def\insertfig#1#2{{{\baselineskip=4pt
\midinsert\centerline{\epsfxsize=\hsize\epsffile{#2}}{{
\centerline{#1}}}\medskip\endinsert}}}
\fi
\input harvmac
%%%%%%%%%%%%%%%%%%%%%%%%%%%%%%%%%%%%%%%%%%%%%%%%%%%%%%%%%%%%%%%%%%%%%%
%
%  UCSD macros to overwrite some of the definitions in harvmac.tex
%  (include after harvmac.tex)
%  last modified 4/92
%
%%%%%%%%%%%%%%%%%%%%%%%%%%%%%%%%%%%%%%%%%%%%%%%%%%%%%%%%%%%%%%%%%%%%%%%
%
% modify the output routine for the little format
%
\ifx\answ\bigans
\else
\output={
  \almostshipout{\leftline{\vbox{\pagebody\makefootline}}}\advancepageno
}
\fi
%
%
% address
%

%
% grant numbers
%

%
% preprint number
%
\def\UCSD#1#2{\noindent#1\hfill #2%
\bigskip\supereject\global\hsize=\hsbody%
\footline={\hss\tenrm\folio\hss}}% restores pagenumbers
%
% abstract
%
\def\abstract#1{\centerline{\bf Abstract}\nobreak\medskip\nobreak\par #1}
%
%
% titlefont
%
%
\edef\tfontsize{ scaled\magstep3}
 \tfontsize  \tfontsize
 \tfontsize \font\titlei=cmmi10 \tfontsize
\font\titleis=cmmi7 \tfontsize \font\titleiss=cmmi5 \tfontsize
\font\titlesy=cmsy10 \tfontsize \font\titlesys=cmsy7 \tfontsize
\font\titlesyss=cmsy5 \tfontsize  \tfontsize
\skewchar\titlei='177 \skewchar\titleis='177 \skewchar\titleiss='177
\skewchar\titlesy='60 \skewchar\titlesys='60 \skewchar\titlesyss='60
%
%\def\titlefont{\def\rm{\fam0\titlerm}% switch to title font
%\textfont0=\titlerm \scriptfont0=\titlerms \scriptscriptfont0=\titlermss
%\textfont1=\titlei \scriptfont1=\titleis \scriptscriptfont1=\titleiss
%\textfont2=\titlesy \scriptfont2=\titlesys \scriptscriptfont2=\titlesyss
%\textfont\itfam=\titleit \def\it{\fam\itfam\titleit}\rm}
%
%
% math symbols
%
%---------------------------------------------------------------------
%
\def\inv{^{\raise.15ex\hbox{${\scriptscriptstyle -}$}\kern-.05em 1}}
  %prime
\def\lbar{{\lower.35ex\hbox{$\mathchar'26$}\mkern-10mu\lambda}} %lambda bar

%
%
% various slashed symbols
%
%
 % slashes a character
\def\dsl{\,\raise.15ex\hbox{/}\mkern-13.5mu D} %this one can be subscripted
\def\delsl{\raise.15ex\hbox{/}\kern-.57em\partial}
\def\Ksl{\hbox{/\kern-.6000em\rm K}}
\def\Asl{\hbox{/\kern-.6500em \rm A}}
\def\Dsl{\hbox{/\kern-.6000em\rm D}} %roman D
\def\Qsl{\hbox{/\kern-.6000em\rm Q}}
\def\gradsl{\hbox{/\kern-.6500em$\nabla$}}
%
% space and backspace in l mode
%
\def\lspace{\ifx\answ\bigans{}\else\qquad\fi}
\def\lbspace{\ifx\answ\bigans{}\else\hskip-.2in\fi} % $$\lbspace...$$
%
%     boxes an equation
%
\def\boxeqn#1{\vcenter{\vbox{\hrule\hbox{\vrule\kern3pt\vbox{\kern3pt
        \hbox{${\displaystyle #1}$}\kern3pt}\kern3pt\vrule}\hrule}}}
%
%     draw a little box (end of proof symbol)
%     e.g. \mbox{.1}{.1}
%
\def\mbox#1#2{\vcenter{\hrule \hbox{\vrule height#2in
\kern#1in \vrule} \hrule}}
%
%
%
%     curly letters
%
   %curly letters

  \def\CO{{\cal O}}

%
%
%
%     derivatives
%
%

%

\def\bar#1{\overline{#1}}

\def\darr#1{\raise1.5ex\hbox{$\leftrightarrow$}\mkern-16.5mu #1}

%
 %pound sterling
%
 %puts a small half in a displayed eqn
\def\frac#1#2{{\textstyle{#1\over #2}}} %puts a small fraction
%in a displayed eqn
%
%
%     various math operators
%
%

\def\GeV{{\rm GeV}}
\def\MeV{{\rm MeV}}

%
%
%
%

%
%       relations
%
\def\ltap{\ \raise.3ex\hbox{$<$\kern-.75em\lower1ex\hbox{$\sim$}}\ }
\def\gtap{\ \raise.3ex\hbox{$>$\kern-.75em\lower1ex\hbox{$\sim$}}\ }
\def\gl{\ \raise.5ex\hbox{$>$}\kern-.8em\lower.5ex\hbox{$<$}\ }
\def\roughly#1{\raise.3ex\hbox{$#1$\kern-.75em\lower1ex\hbox{$\sim$}}}
%
%
%       This defines et al., i.e., e.g., cf., etc.

\def\etal{\hbox{\it et al.}}

\def\np#1#2#3{{Nucl. Phys. } B{#1} (#2) #3}
\def\pl#1#2#3{{Phys. Lett. } {#1}B (#2) #3}
\def\prl#1#2#3{{Phys. Rev. Lett. } {#1} (#2) #3}
\def\physrev#1#2#3{{Phys. Rev. } {#1} (#2) #3}

\relax

\def\hbar{\bar h_Q}

\def\qsl{\hbox{/\kern-.5600em {$q$}}}
\def\ksl{\hbox{/\kern-.5600em {$k$}}}

\def\({\left(}
\def\){\right)}

\def\lamtwo{\lambda_2}

\def\OMIT#1{}
\def\frac#1#2{{#1\over#2}}

\def\etal{{\it et. al.}}

\def\vbc{\vert V_{bc}\vert}
\def\asoverpi#1{ {\alpha_s(#1)\over \pi} }

\hbadness=10000

\noblackbox
\vskip 1.in
\centerline{{\titlefont{Extracting $\vbc$, $m_c$ and $m_b$
from}}}
\medskip
\centerline{{\titlefont{Inclusive $D$ and $B$ Decays
}}}
\vskip .5in
\centerline{Michael Luke${}^{ab}$ and
Martin J.~Savage${}^{bc}$}
\medskip
{\it{
\centerline{a) Department of Physics, University of Toronto, Toronto,
Canada M5S 1A7}
\centerline{b) Department of Physics, University of California, San
Diego, La Jolla CA 92037}
\centerline{c) Department of Physics, Carnegie Mellon University,
Pittsburgh PA 15213}}}

\vskip .2in

\abstract{
Using recent results for nonperturbative contributions to the $B$ and
$D$ meson inclusive semileptonic widths, a model independent
extraction of $\vbc$, $m_c$ and $m_b$ is made from the
experimentally measured $B$ and $D$ lifetimes and semileptonic branching
ratios.  Constraining the parameters of the HQET at $\CO(1/m_Q^2)$
by the $D$ semileptonic width,
$\vbc$ is found to lie in the range $.040<\vbc< 0.057$.
The $c$ and $b$ quark masses are not well constrained due
to uncertainty in the relevant scale of $\alpha_s$.
These results assume
the validity of perturbative QCD at the low scales relevant to semileptonic
charm decay.  Without making this assumption,
somewhat less stringent bounds on $V_{bc}$ from $B$ decay alone
may be obtained.}

\vfill
\UCSD{\vbox{
\hbox{UCSD/PTH 93-25}
\hbox{UTPT 93-21}
\hbox{CMU-HEP 93-13}
\hbox{DOE-ER/40682-37}}
}{Revised version, October 1993}
\eject

There has been much recent interest in the application of the
techniques of the heavy quark effective theory (HQET)
\ref\volshi{M.~B.~Voloshin and M.~A.~Shifman, Yad. Phys.
45 (1987) 463 [Sov. J. Nucl. Phys. 45 (1987) 292];
Yad. Fiz. 47 (1988) 801 [Sov. J. Nucl. Phys. 47 (1988) 511].}--%
\nref\iw{N.~Isgur and M.~B.~Wise, \pl{232}{1989}{113}, \pl{237}{1990}{527}.}%
\nref\hg{H.~Georgi, \pl{240}{1990}{247}.}%
\nref\bgrin{B.~Grinstein, \np{339}{1990}{253}.}%
\ref\eh{E.~Eichten and B.~Hill, \pl{243}{1990}{427}.}
to inclusive decays of hadrons containing a single c or b quark
\ref\cgg{J.~Chay, H.~Georgi and B.~Grinstein, \pl{247}{1990}{399}. }--%
\nref\bigietala{I.I.~Bigi, N.~G.~Uraltsev and A.~I.~Vainshtein,
\pl{293}{1992}{430}.}%
\nref\bigietalb{I.~I.~Bigi, B.~Blok, M.~Shifman, N.~G.~Uraltsev and
A.~Vainshtein,
TPI-MINN-92/67-T (1992); \prl{71}{1993}{496}.}%
\nref\bloketal{B.~Blok, L. Koyrakh, M.~Shifman and A.~Vainshtein,
NSF-ITP-93-68, hep-ph/9307247.}%a
\nref\manwise{A.~V.~Manohar and M.~B.~Wise, UCSD-PTH 93-14, hep-ph/9308246.}%
\nref\mannel{T.~Mannel, IKDA 93/16, hep-ph/9308262.}%
\ref\fls{A.~F.~Falk, M.~Luke and M.~J.~Savage, UCSD-PTH 93-23.}.
As first made explicit in \cgg, the differential decay rate for
an inclusive factorisable process such as $B\rightarrow X_c e\nu_e$ or
$B\rightarrow X_s \gamma$ may written, using an operator product
expansion, in terms of the matrix elements of local operators
between $B$ mesons.
Using the techniques of HQET, it was shown in \cgg\ that the leading
term of the
expansion in $1/m_b$ reproduces the parton model, and that
the $\CO(1/m_b)$ corrections vanish by the equations of motion for
the heavy quark.
More recently,
the $\CO(1/m_b^2)$ corrections to the parton model have
been calculated for semileptonic decays
\bigietala--\mannel, as well
as for the rare decays $B\rightarrow X_s\gamma$ \bigietalb\fls\ and
$B\rightarrow X_s e^+ e^-$ \fls.
A crucial observation which emerges from the operator product expansion is
that the semileptonic width of a heavy meson is proportional to the fifth
power of the heavy quark mass, not the meson mass.  The
heavy quark mass $m_q$ is a well-defined quantity in HQET, and is defined
such that the residual mass term in the effective heavy quark Lagrangian
vanishes (for a detailed discussion of the definition of $m_q$, see
\ref\fln{A.~F.~Falk, M.~Luke and M.~Neubert, \np{388}{1992}{363}.}).

In addition to these nonperturbative corrections, there are calculable
$\CO\big(\alpha_s(m_q)\big)$ corrections to the free quark decay picture
coming from real and virtual gluon emission
\ref\cabmai{N.~Cabibbo and L.~Maiani, \pl{79}{1978}{109}.}.
In this letter, the complete expression to $\CO\(\alpha_s(m_q),
\ 1/m_q^2\)$ for the semileptonic width of a heavy meson is
used to extract the c
quark mass from the measured $D^\pm$ lifetime and semileptonic branching
ratio.  From this, the b quark mass is determined and the
$B$ lifetime predicted in terms of the KM matrix element $\vbc$.  Comparing
this
with the experimentally measured lifetime allows $\vbc$ to be determined.

Throughout this letter the mass of a heavy meson will be denoted by
$M_Q$ and the mass of the corresponding quark by $m_q$. The meson and quark
masses are related by
\ref\fn{A.~F.~Falk and M.~Neubert, \physrev{D47}{1993}{2965}.}
\eqn\mesmass{M_Q=m_q+\overline{\Lambda}- {\lambda_1+3\lambda_2\over
2m_q}+\CO\({1\over m_q^2}\).} The parameter $\bar\Lambda$ in
\mesmass\ may be interpreted as the energy of the light degrees of freedom
of the meson.  It corresponds to the constituent mass of the light quark in
the meson, and so is expected to be of order a few hundred MeV. {}From
consideration of QCD mass inequalities analogous to those used in the light
quark sector to prove that
$m_\rho>m_\pi$, Guralnik and Manohar \ref\zach{Z. Guralnik and A.V. Manohar,
\pl{302}{1993}{103}.} have shown the existence of a rigorous lower bound
\eqn\gmbound{\bar\Lambda>237\,\GeV.}
We will also take $\bar\Lambda<800\,\MeV$ in this work when quoting
limits on $m_c$, $m_b$ and $V_{bc}$.  This bound is consistent with
the QCD sum rules estimate \ref\mnsumrules{M.~Neubert,
\physrev{D46}{1992}{1076}.}\ (see also the second
reference in \ref\vbcexcl{M.~Neubert, \pl{264}{1991}{455};
SLAC-PUB-6263 (1993)
(to appear in Physics Reports).})
\eqn\lbsr{{\bar\Lambda}^{\rm s.r.}=570 \pm 70\, \MeV}
and is also consistent with the quark model, in which $\bar\Lambda$
is expected to be a few hundred MeV.

$\lambda_1$ and $\lambda_2$ are defined in terms of
the expectation values in the heavy quark effective theory,
\eqn\lamdef{\eqalign{&<H_Q|\bar h (iD)^2 h|H_Q>=2 M_Q \lambda_1\cr
&<H_Q|\bar h \({{\rm i}\over 2}\)\sigma^{\mu\nu}G_{\mu\nu}\  h|H_Q> = 6
M_Q\lambda_2(\mu)}} where $H_Q$ is the pseudoscalar heavy meson ($B$ or $D$)
and $h$ is the heavy quark field in the effective theory.
$\lambda_2$ parameterises the effects of the chromomagnetic moment operator,
and may be extracted from the $B^*-B$ mass splitting
\eqn\lamtwo{\lambda_2(m_b)={m_b\over 8}(M_{B^*}-M_B) \simeq
{1\over4}(M_{B^*}^2-M_B^2)=0.12\,\GeV^2} (corresponding to
$\lambda_2(m_c)=0.10\,\GeV^2$ \ref\flgeh{ A.~F.~Falk, B.~Grinstein and
M.~Luke, \np{357}{1991}{185}\semi E.~Eichten and B.~Hill,
\pl{243}{1990}{427}}).
$\lambda_1$ parameterises the kinetic energy of the b quark inside the
hadron.  Since it contributes equally to $m_B$ and $m_{B^*}$,
$\lambda_1$ cannot be extracted from the meson masses.

At $\CO(1/m_q^2)$, the nonperturbative corrections to the parton
model are also parameterised in terms of $\lambda_1$ and
$\lambda_2$.  Combining the results from
\bigietala--\mannel\ with the
$\alpha_s$ corrections to the free quark model \cabmai\ gives
the complete expression for the semileptonic $D$ width to
$\CO\big(1/m_c^2,\ \alpha_s(\mu_c)\big)$
\eqn\semic{\eqalign{\Gamma(D\rightarrow e^+\nu_e X_q)=
{G_F^2m_c^5\over 192\pi^3} &\left\{
|V_{cs}|^2\left[
\left(1-{2\alpha_s(\mu_c)\over 3\pi}g\Big({m_s\over m_c}\Big)
+{\lambda_1\over 2m_c^2}\) f_1\Big({m_s\over m_c}
\Big)\right.\right.\cr
&\qquad\qquad\left.\left.
-{9\lambda_2\over 2m_c^2} f_2\Big({m_s\over m_c}\Big) \right)\right]\cr
&+\left.|V_{cd}|^2\left[
1-{2\alpha_s(\mu_c)\over 3\pi}g(0)
+{\lambda_1-9\lambda_2\over 2m_c^2}\right]\right\}\cr
&+\CO\({1\over m_c^3}, \alpha_s(\mu_c)^2,
{\alpha_s(\mu_c)\over m_c^2}\).}}
The functions $f_{1,2}$
arise from the finite mass of the quark in the final state,
\eqn\phase{\eqalign{
&f_1(x) = 1-8x^2+8x^6-x^8-24x^4\log x \cr
&f_2(x) = 1-{8\over 3}x^2-8x^4+8x^6+{5\over 3}x^8+8x^4\log x,}}
while the function $g(x)$ arising from one-gluon graphs is tabulated in
Ref.~\cabmai\   (we note that $g(0)=3.62$, $g(.2/1.6)=3.15$ and
$g(1.6/4.8)= 2.40$).  We also take
$m_s=200\pm 80\,\MeV$ and $m_d\simeq 0$.

The choice of the appropriate scale $\mu_c$ in \semic\ is
a potentially troublesome one.  The scale is set roughly
by $m_c$; however, since the final state hadron typically carries
off only a fraction of the available energy, the
appropriate scale may be somewhat lower.
Formally
this choice of scale is a higher order effect; however, in practice
there is a considerable difference between, for example, $\alpha_s(m_c)$
and $\alpha_s(m_c/3)$.  In particular, at scales much less
than $m_c$, QCD perturbation theory
cannot be trusted.  Since we have not done the two-loop calculation,
we cannot resolve this scale ambiguity.
In this letter we will take $\mu_c\simeq m_c$, and
the sensitivity of the results to higher order terms in $\alpha_s$ will
be estimated by formally
varying $\mu_c$ between $m_c$ and $m_c/3$.  Taking
$\bar\Lambda^{(4)}_{{\rm QCD}}=250\,\MeV$ and
$\bar\Lambda^{(3)}_{{\rm QCD}}=300\,\MeV$, and using the two-loop
expression for $\alpha_s(\mu)$ gives
$\alpha_s(m_c)=0.09\,\pi$ and $\alpha_s(m_c/3)\simeq 0.4\,\pi$.
We stress that we are not claiming that QCD perturbation theory
is valid at $\mu\simeq m_c/3$, but rather that
if perturbation theory is meaningful for $D$ decays,
we expect this approach to give a reasonable estimate of the uncertainty
due to higher order QCD
corrections.\footnote{$^\dagger$}{Note that the energies encountered in this
problem are similiar to those encountered in hadronic $\tau$ decays,
for which the corrections to $\CO\(\alpha_s(m_\tau)^4\)$ have
been calculated, and for which the perturbation series appears to be
converging \ref\bnp{E.~Braaten, S.~Narison and A.~Pich,
\np{373}{1992}{581}, and references therein.}.  }
Fortunately, the determination of
$V_{bc}$ is relatively insensitive to the uncertainty in
$m_c$ extracted from the $D$ width.

{}From the observed semileptonic branching ratio, lifetime and mass of the
$D^\pm$ \ref\pdg{Particle Data Group, \physrev{D45}{1993}{1}.}
\eqn\dstuff{\eqalign{
&{\rm Br}(D^\pm\rightarrow e^\pm + {\rm anything}) =  (17.2\pm 1.9)\%\cr
&\tau_{D^\pm} = 10.66\pm 0.23 \times 10^{-13}\ {\rm s}\cr
&M_{D^\pm}=1869.3\pm 0.3\, \MeV}}
it is a simple matter to extract $m_c$ and $\lambda_1$ from
Eqs.~\semic\ and \mesmass\ for any
value of $\bar\Lambda$. (We have checked that the $D^0$ gives results
consistent with the $D^\pm$; the $D^\pm$ meson is used, as
the uncertainty in its semileptonic branching fraction is
somewhat smaller than for the $D^0$.) The results are
shown in \fig\firstfig{$\lambda_1$, $m_c$ and $m_b$ as functions of
$\bar\Lambda$.  In each graph,
the vertical line corresponds to the Guralnik-Manohar bound
$\bar\Lambda>237\,\GeV$.  The shaded regions
correspond to the experimental error in the input parameters
and, for the broader band, the variation in $\mu_c$.}
(a) and (b).  In each figure the narrow band corresponds to
$\mu_c=m_c$, while the broader band correspond to
$m_c/3<\mu_c<m_c$.  The uncertainty corresponding
to $1\sigma$ errors on all experimental input parameters is included
in the width of each band.
{}From \firstfig, the Guralnik-Manohar bound \gmbound\ on $\bar\Lambda$
translates to limit
\eqn\lamonelim{\lambda_1> -0.5\,\GeV\ (1\,\sigma).}
Because of the uncertainty in $\mu_c$, the c quark mass is not
well constrained from the data, although taking $\bar\Lambda\ltap 800\,\MeV$
suggests $m_c\gtap 1460\,\MeV$.  Since
the upper limit of the error bar in \firstfig (a) is both very sensitive to
the value of $\alpha_s$ at low scales and corresponds to
$m_c>m_D$, a useful upper bound on $m_c$ cannot be extracted.

For $B$ mesons, formulas analogous to \mesmass\ and \semic\ hold,
with the replacements $M_D\rightarrow M_B$, $m_c\rightarrow
m_b$, $V_{cs}\rightarrow V_{bc}$, $V_{cd}\rightarrow V_{bu}$ and
$m_s\rightarrow m_c$.  Since $\lambda_1$ is now determined from $D$
decays for
each value of $\bar\Lambda$,  the b quark
mass $m_b$ may be solved for as a function only of $\bar\Lambda$.
Using $M_B = 5.278\pm 0.002\,\GeV$ gives the results shown in
\firstfig(c).   Since $M_B$ is less sensitive to $\lambda_1$
than $M_D$, it is not surprising that $m_b$ is roughly a
linear function of $\bar\Lambda$, with smaller error
bars than $m_c$.  $m_b$ is found to lie in the range
$5140\,\MeV\gtap m_b \gtap 4600\,\MeV$.

Given $m_b$ and $\lambda_1$, the semileptonic width of the $B$ may now
be predicted, allowing
the KM matrix element $\vbc$ to be determined as a function of $\bar\Lambda$.
The inclusive branching ratio of the
$B$ to electrons is measured to be \pdg
\eqn\semib{{\rm Br}(B\rightarrow e + {\rm anything}) = 10.7\pm 0.5 \%}
(where the charge of the $B$ is not determined).
Averaging the recent measurements of the $B$ lifetime by the DELPHI
collaboration
\ref\delphi{P. Abreu, \etal, (DELPHI Collaboration), CERN-PPE/93-80 (1993).}
and by the CDF collaboration
\ref\cdf{M.L. Mangano (CDF Collaboration), FERMILAB-Pub-93/139-E (1993).}
gives a $B$ lifetime of
\eqn\btimecdf{\tau_B = 1.32\pm 0.09 \times 10^{-12}\ {\rm s}}
(statistical and systematic errors added in quadrature).
This gives
$\vbc$ as a function
of $\bar\Lambda$, as shown in
\fig\vbcfig{The weak mixing angle $\vbc$ as a
function of $\bar\Lambda$ for $\tau_B = 1.32\pm 0.09$ ps.
The vertical line corresponds to the constraint imposed by the
bound on $\bar\Lambda$.
The error bar on the left indicates the current value of $\vbc$
extracted from the exclusive decay
$B\rightarrow D^* e\bar\nu_e$
and its associated uncertainty.  The shaded
region corresponds to the experimental error
in the input parameters and, for the broader band,
the variations in $\mu_c$ and $\mu_b$.}.
The two bands in the figure correspond
to the two choices of scale $\mu_b=m_b,\,\mu_c=m_c$ and
$m_c/3<\mu_c<m_c$, $m_b/3<\mu_b<m_b$ (with
$\mu_c/\mu_b$ fixed).
Over the range of values for $\bar\Lambda$ shown in \vbcfig,
$\vbc$ is found to lie in the range $0.040 < \vbc< 0.057$.
The lower bound is set by
the rigorous lower bound on $\bar\Lambda$, while the upper corresponds to
$\bar\Lambda=800\,\MeV$.
This extraction is consistent with the current value of
$\vbc\left( 1.32 \ {\rm ps}/\tau_B\right)^{1/2} = 0.051\pm 0.008 $
extracted from the exclusive decay $B\rightarrow D^*e\bar\nu_e$
using HQET \vbcexcl.
We note that choosing
a lower scale for $\mu_c$ and $\mu_b$ tends to raise the preferred
values of $\vbc$.

For a given value of $\bar\Lambda$ the effects of higher order
terms in $1/m_q$ and $\alpha_s$ on this extraction of $\vbc$ may
be estimated.  These terms may be parametrized by
$\delta_{b,c}$ and $\epsilon_{b,c}$, defined by
\eqn\errormesmass{\eqalign{M_D=&m_c\(1+{{\bar\Lambda}\over m_c}-
{\lambda_1+3\lambda_2\over 2m_c^2}+\delta_c\)\cr
M_B=&m_b\(1+{{\bar\Lambda}\over m_b}-
{\lambda_1+3\lambda_2\over 2m_b^2}+\delta_b\)
}}
and
\eqn\errorsemic{\eqalign{\Gamma(D\rightarrow X_q e^+\nu_e)
&\propto m_c^5(1+\epsilon_c)\cr
\Gamma(B\rightarrow X_q e^+\nu_e)
&\propto m_b^5(1+\epsilon_b).}}
They correspond to a change in $\vbc$ of
\eqn\vbcerror{{\Delta \vbc\over \vbc}\simeq {1\over 2}\({m_c\over
m_b}(\epsilon_c+5\delta_c)-(\epsilon_b+5\delta_b)\)
\simeq {1\over 6}\big(\epsilon_c-3\epsilon_b+5(\delta_c-3\delta_b)\big).}
The leading contributions to $\delta_c$ are of order $(\bar\Lambda/m_c)^3$ and
$\asoverpi{m_c}\lambda_2/m_c^2$\footnote{$^\dagger$}{The operator
$\bar h (iD)^2 h$ receives
no strong interaction corrections to its coefficient, by reparameterisation
invariance
\ref\lm{M. Luke and A. V. Manohar, \pl{286}{1992}{348}.}.},
which we estimate to be at most a few percent, while $\delta_b$ is
expected to be at least an order of magnitude smaller.
We have attempted to take the leading contributions to $\epsilon_{c,b}$
arising from higher order QCD corrections into account by varying the
scales $\mu_{c,b}$ between $m_{b,c}$ and $m_{b,c}/3$.  From
\vbcerror, we can understand why our results were relatively insensitive
to the choice of scale: an uncertainty in the charm width only gives
$\sim 1/6$ this uncertainty in $\vbc$.  Furthermore, the uncertainty in scale
tends to cancel between $\epsilon_c$ and $\epsilon_b$.

Since the extraction of $\lambda_1$ as a function of $\bar\Lambda$
depends on the validity of perturbation theory for QCD at a
low ($\mu\ltap m_c$) scale, it is instructive to compare this extraction
of $\vbc$ with that obtained by ignoring the
$D$ decay data altogether and simply varying $\lambda_1$ between
$\pm 1\,\GeV^2$ for each given value of $\bar\Lambda$.  This amounts to
working at $\CO\(1/m_q\)$ in the heavy quark expansion with a reliable
estimate of the uncertainty from $1/m_q^2$ terms.  As before,
the scale $\mu_b$ is also varied between
$\mu_b=m_b/3$ and $\mu_b=m_b$.  The results are shown in \fig\nocharm{
The weak mixing angle $\vbc$ as a function of $\bar\Lambda$ for
$\tau_B=1.32\pm 0.09$ ps, without imposing constraints on
$\lambda_1$ from $D$
decay.  The vertical line corresponds to the lower bound
on $\bar\Lambda$, and the shaded regions correspond to the experimental
error in the input parameters, varying $\lambda_1$ from $-1\,\GeV$ to
$1\,\GeV$ and $\mu_b$ from $m_b$ to $m_b/3$.
The error bar on the left indicates the current value of $\vbc$
extracted from the exclusive decay
$B\rightarrow D^* e\bar\nu_e$
and its associated uncertainty.}.
The shaded area on each graph
corresponds to both the experimental errors on the input parameters
and the variations in $\lambda_1$ and $\mu_b$.  As expected, the
uncertainty is $\vbc$ is somewhat larger than when the $D$ decay
data are included, particularly at large values of $\bar\Lambda$;
however, $\vbc$ is still well constrained since the prediction
for the $B$ width is relatively insensitive to $\lambda_1$.
For $237\,\MeV<\bar\Lambda <800\,\MeV$, $\vbc$ is
found to lie in the range $0.04<\vbc<0.06$.

It is encouraging that the extraction of $\vbc$ from inclusive
decays is consistent with that
obtained from exclusive decays over the full allowed range of $\bar\Lambda$.
Furthermore, the extraction of $\vbc$ from inclusive decays has
some theoretical and experimental advantages over exclusive
decays.   No exclusive hadronic final state needs to be identified,
and the extrapolation of form factors to zero recoil is not required.
The method is limited by the uncertainties in $\bar\Lambda$ and
the relevant hadronic scales $\mu_b$ and $\mu_c$.
Additional input, such as a lattice measurement of
$\bar\Lambda$ or $\lambda_1$, or a two-loop calculation
of the semileptonic decay rate, would further
constrain $\vbc$. For example,
{}from \firstfig (a), the QCD sum rules estimate \lbsr\ of $\bar\Lambda
=570 \pm 70\, \MeV$ corresponds to
$0.1\,\GeV^2<\lambda_1\ltap 1.5\,\GeV^2$ and $0.044<\vbc<0.054$
(although the upper bound on $\lambda_1$ should not be taken too seriously,
since both the $1/m_c$ expansion and QCD perturbation theory
are unreliable at this point).
On the other hand, the nonrelativistic quark model suggests that
$\bar\Lambda\sim 350\,\MeV$, the mass of the light constituent
quark,  and that $\lambda_1\sim -(350\,\MeV)^2\sim -0.1\GeV^2$
(note that $\lambda_1<0$ corresponds to a positive mass shift
from the kinetic energy of the heavy quark).  From \firstfig (a),
these values of $\lambda_1$ and $\bar\Lambda$
are consistent with one another, and are consistent with
somewhat lower values of $\vbc$.

\bigskip

We thank M.~B.~Wise for useful discussions, particularly concerning
the relevant scales $\mu_c$ and $\mu_b$ and the possibilities
of extracting $V_{bc}$ without using the $D$ decay data. We are also
grateful to A.~Falk, B.~Holdom, A.~Manohar and
L.~Wolfenstein for useful comments and criticisms.
MJS acknowledges the support of a
Superconducting Supercollider National Fellowship from the Texas
National Research Laboratory Commission under grant  FCFY9219.
This research was supported in part by TNRLC grant RGFY93-206
and by the Department of Energy under contracts DE--FG03--90ER40546
(UC San Diego) and DE--FG02--91ER40682 (CMU).

\listrefs
\listfigs
\vfill\eject
\insertfig{Figure 1 (a)}{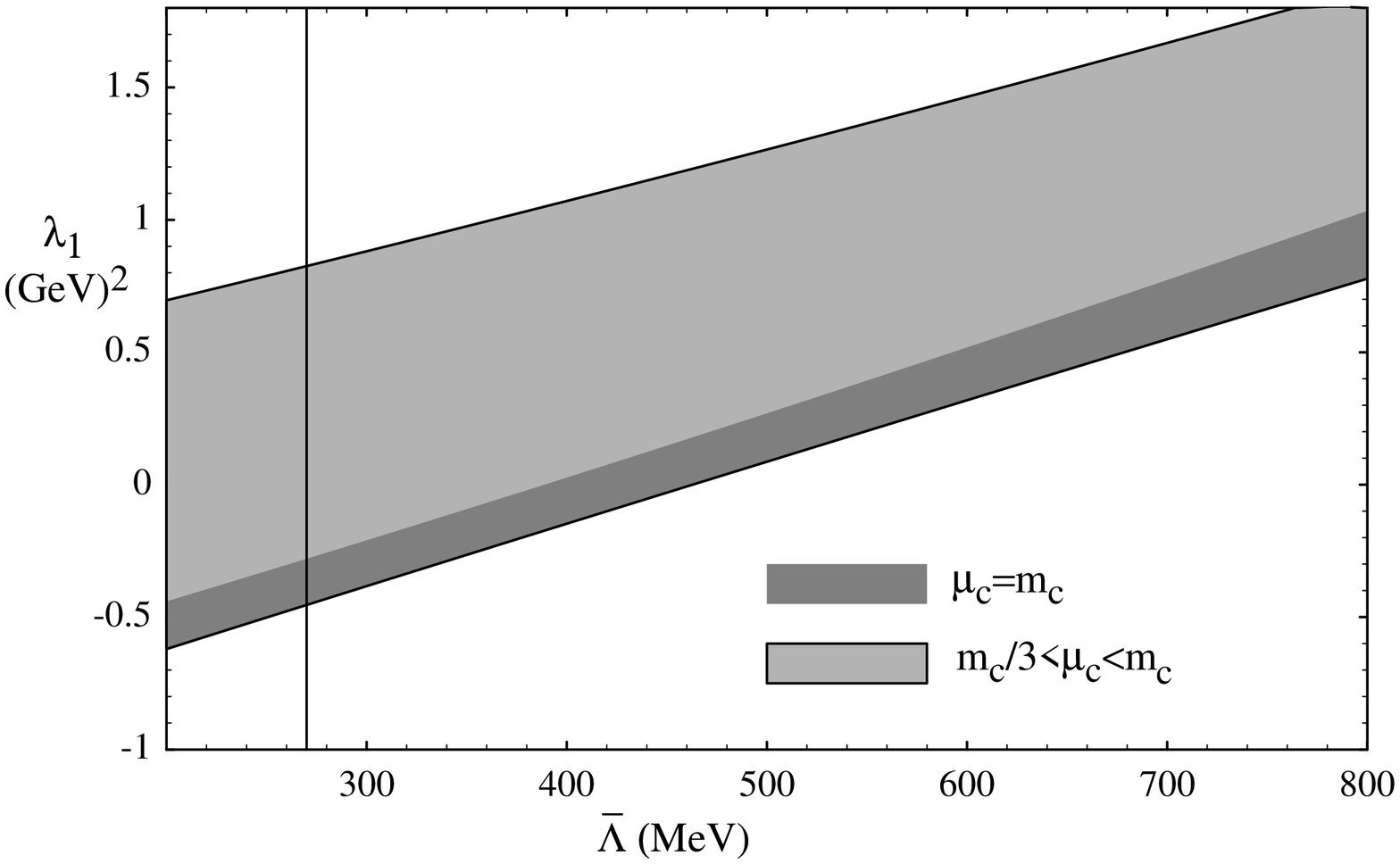}
\insertfig{Figure 1 (b)}{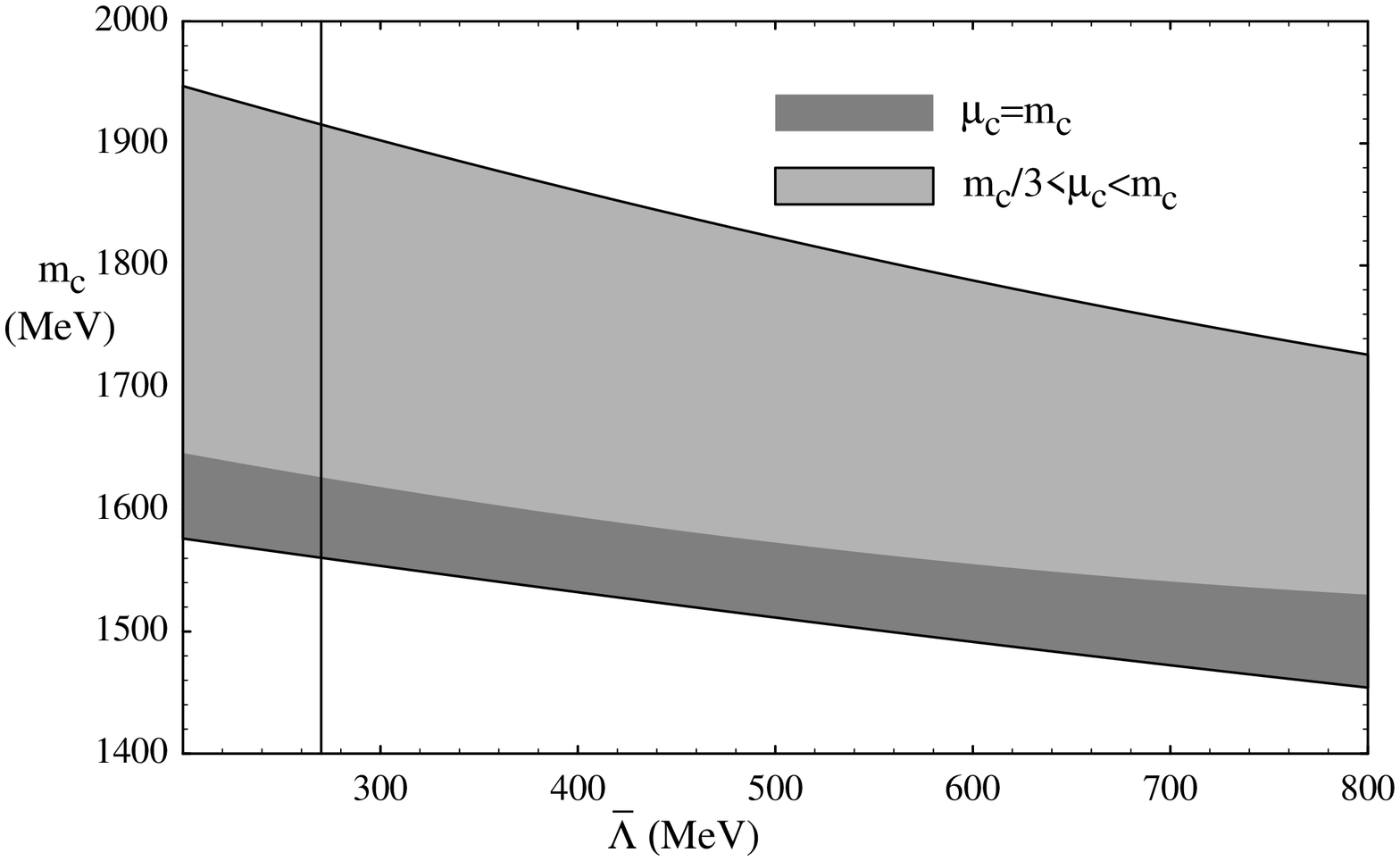}
\insertfig{Figure 1 (c)}{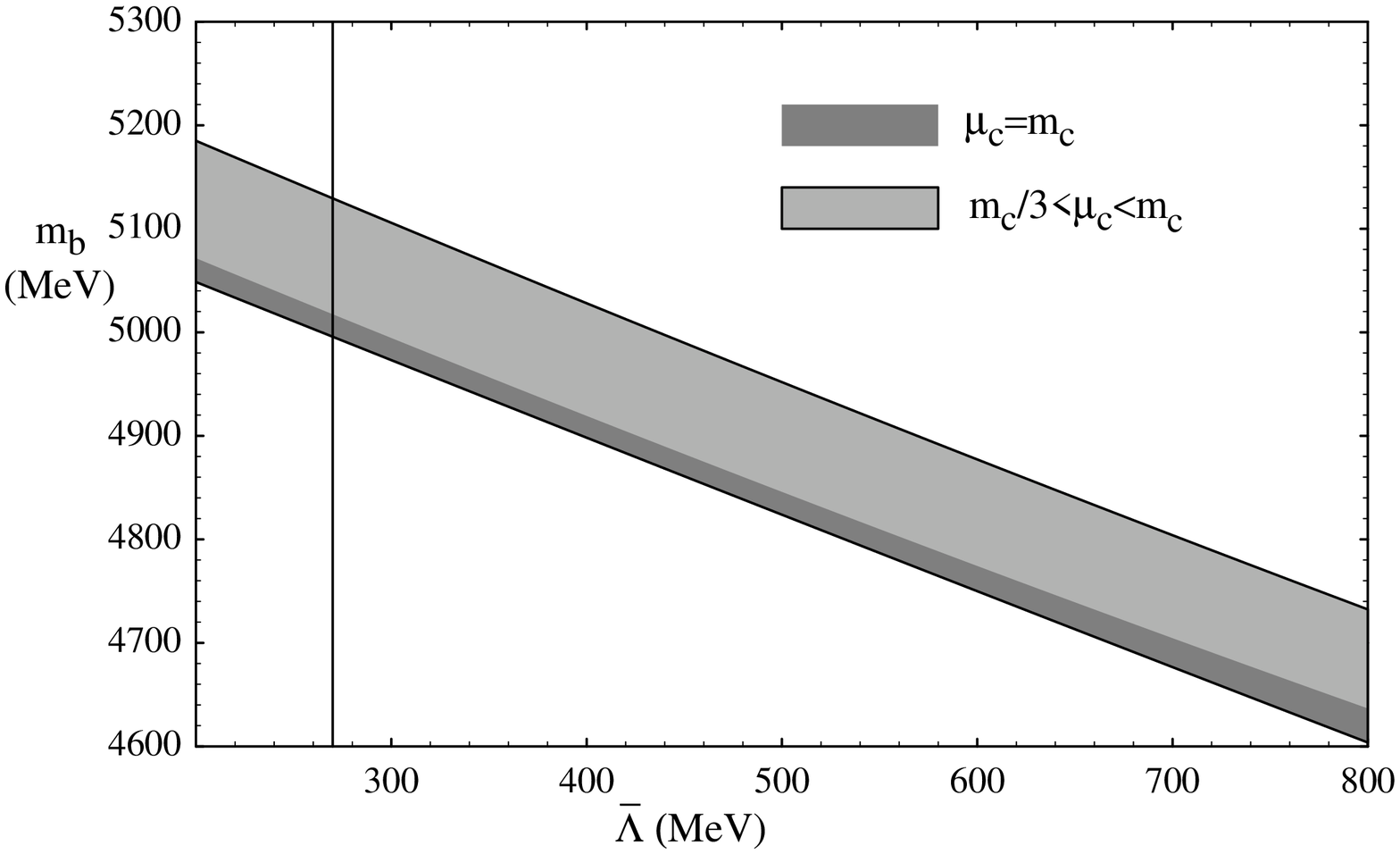}
\insertfig{Figure 2}{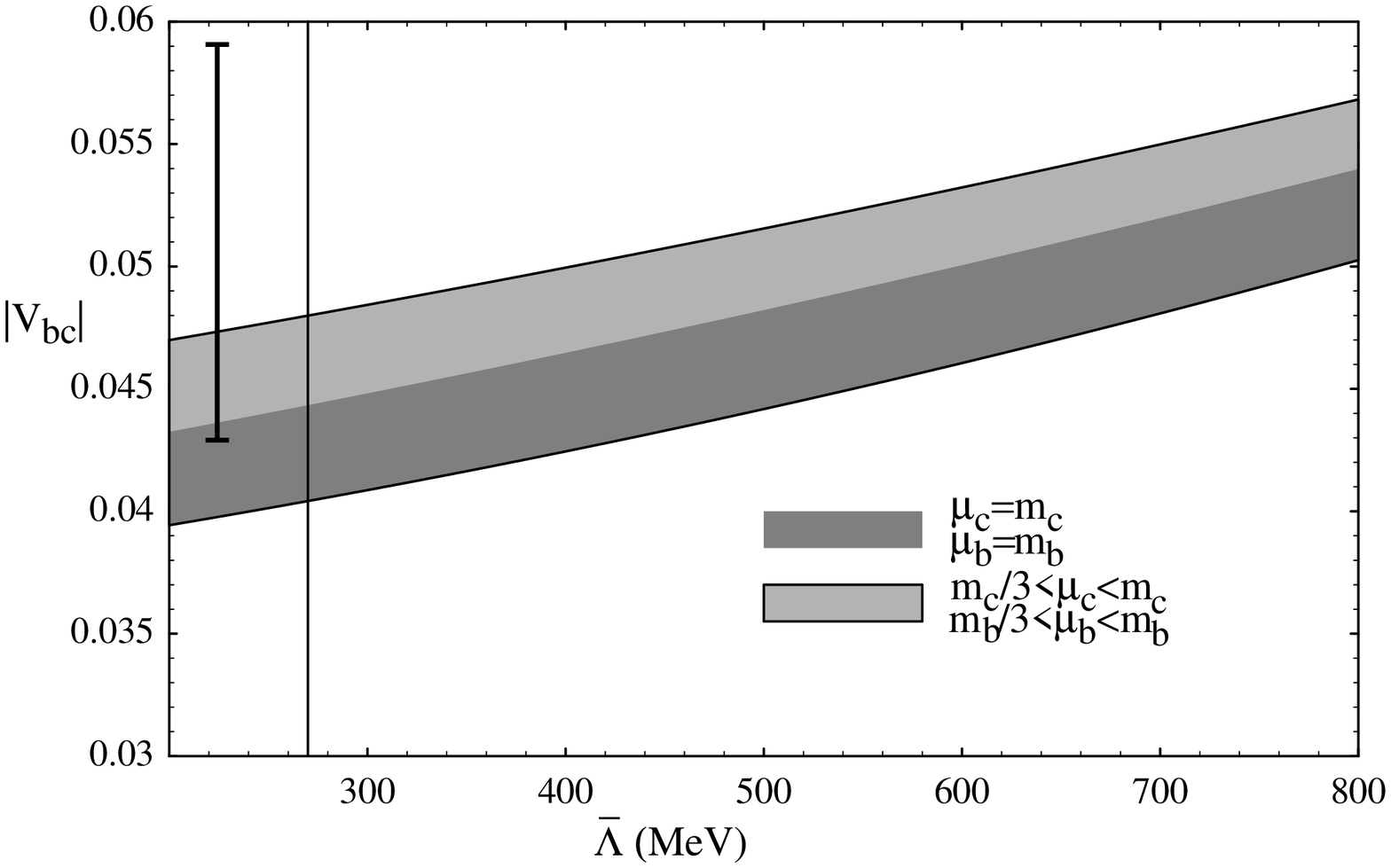}
\insertfig{Figure 3}{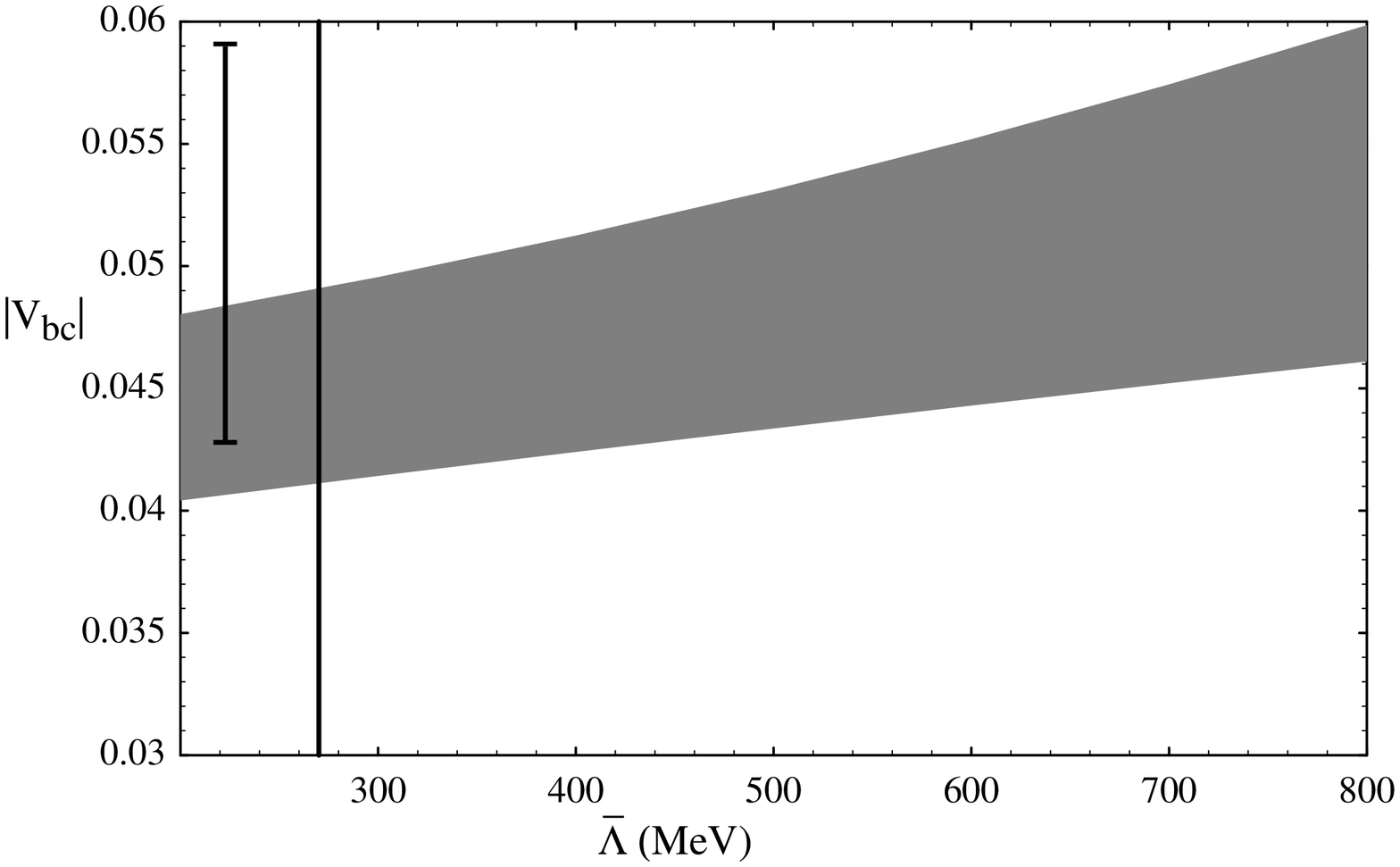}
\vfill\eject
\bye